\documentclass{aa} 

\usepackage{graphicx}
\usepackage{txfonts}

\begin{document}

   \title{One size does not fit all: Evidence for a range of mixing efficiencies in stellar evolution calculations}
    \titlerunning{Convective cores and chemical mixing}

\author{C. Johnston\inst{1,2}}

\institute{Department of Astrophysics, IMAPP, Radboud University Nijmegen, P. O. Box 9010, 6500 GL Nijmegen, the Netherlands, \email{cole.johnston@ru.nl}
    \and 
    Institute of Astronomy, KU Leuven, Celestijnenlaan 200D, 3001, Leuven, Belgium  }

   \date{Received 14 April 2021; accepted 13 July 2021}

 
  \abstract
   {Internal chemical mixing in intermediate- and high-mass stars represents an 
    immense uncertainty in stellar evolution models. In addition to extending the 
    main-sequence lifetime, chemical mixing also appreciably increases the mass 
    of the stellar core. Several studies have made attempts to calibrate the 
    efficiency of different convective boundary mixing mechanisms, with 
    sometimes seemingly conflicting results.  }
   {We aim to demonstrate that stellar models regularly under-predict the masses of convective
    stellar cores. }
   {We gather convective core mass and fractional core hydrogen content inferences from 
    numerous independent binary and asteroseismic studies, and compare them to stellar 
    evolution models computed with the MESA stellar evolution code.}
   {We demonstrate that core mass inferences from the literature are ubiquitously more 
    massive than predicted by stellar evolution models without or with little convective 
    boundary mixing.}
   {Independent of the form of internal mixing, stellar models
    require an efficient mixing mechanism that produces more massive cores throughout the 
    main sequence to reproduce high-precision observations. This has implications for the 
    post-main sequence evolution of all stars which have a well developed convective core 
    on the main sequence.}

   \keywords{Asteroseismology -
                Stars: interiors -
                Stars: evolution -
                Stars: oscillations (including pulsations) -
                Stars: binaries: eclipsing 
               }

   \maketitle
%

\section{Introduction}

The treatment of convection and of convective boundary mixing has 
major implications for the results of stellar evolutionary models
\citep{Maeder2009,Kippenhahn2012}. As it is inherently a 3D process, 
the 1D descriptions of convection required for modern stellar structure
and evolution calculations, often achieved using the Mixing Length 
Theory \citep[MLT, ][]{BohmVitense1958} or variations thereof, have 
shortcomings. Although MLT does well in approximating the deep 
interiors of a convective region, its limitations become apparent at 
the boundaries of convective regions. Namely, MLT does not capture 
the scope of processes that occur over different distance scales at 
the interface of a turbulent convective zone and a stably stratified 
radiative zone. In such transition regions, phenomena including 
convective penetration, overshooting, turbulent entrainment, bulk
shear mixing, and the generation of internal gravity waves which 
induce mixing in the adjacent radiative region may occur \citep{Renzini1987,
Arnett2019}. Additionally, 1D simplifications of convection result 
in a single, hard convective boundary (based either on the Schwarzschild 
or Ledoux criterion), rather than the dynamical, fluctuating boundary 
that is often observed in 2- and 3D hydrodynamical simulations 
\citep{Meakin2007,Arnett2015,Cristini2017,Higl2021}. As a result of 
ignoring the non-local extensions of convection, simplified 1D descriptions 
of convection underestimate the transport of chemical species both at the 
convective boundary and throughout the adjacent radiative region. More 
specifically, in the absence of some form of explicitly included extra 
chemical mixing, 1D stellar models with a convective nuclear burning core 
underestimate the transport of fresh chemical species from the envelope 
into the core where they can participate in burning. This implies that models 
which do not include extra mixing underestimate the mass of the convective 
core at any time along a star's evolution when the core is convective 
\citep{Bressan1981,Stothers1981,Bertelli1985,Maeder1987}.

To account for the missing chemical mixing induced at convective boundaries, 
1D stellar evolution models introduce ad-hoc parameterised mixing profiles
that are stitched to convective regions. However, each parameterised mechanism 
only serves to address one aspect of the many processes occurring at, or beyond, 
a convective boundary \citep{Meakin2007,Arnett2015,Rogers2017}. Furthermore, 
all of these parameterised descriptions have at least one free parameter that 
scales the `efficiency' of the chemical mixing induced by that mechanism. 
Numerous studies have attempted to observationally calibrate the efficiency 
of different mixing mechanisms both at the convective boundary and in the 
radiative envelope, however, these studies often report conflicting efficiencies 
for the same mechanisms. The focus of this paper is to discuss the complexities 
of observationally constraining the efficiency of mixing mechanisms in stellar
models, to introduce convective core mass as a robust point of comparison 
for the implementation of chemical mixing in 1D stellar models, and to 
consider the consequences of a range of inferred mixing efficiencies in model
calculations.

\section{Chemical mixing in the literature}

A complete chemical mixing profile should include the effects of 
entrainment, penetration, overshooting, semi-convection, as well 
as mixing induced by rotational instabilities and internal gravity 
waves \citep{Langer2012,Maeder2009,Hirschi2014,Arnett2015,Rogers2017,
Salaris2017}. Unfortunately, discussion of chemical mixing mechanisms 
in the literature suffers from a problem of mixed identities. Numerous 
studies have attempted to constrain the impact of chemical mixing in 
different forms, be it any of the mechanisms previously mentioned. 
However, more often than not, these studies focus on constraining the
efficiency of a single mixing mechanism, fixing the efficiency of the 
other mechanisms or ignoring them outright. In the process of doing 
this, one effectively enforces that the mixing mechanism under investigation 
in practice becomes a proxy for all of the chemical transport present. 
Thus, when comparing results of different studies, one must account
for both the physical mechanisms that were explicitly included, as well 
as the absence of effects from any physical mechanisms that were not 
included.

Given modern observational quantities, their associated precision, and 
underlying model assumptions and degeneracies, we are unable to disentangle 
the contributions of individual mixing mechanisms to the point of robustly 
calibrating multiple mixing mechanisms in a single target \citep{Valle2017,
Valle2018,Constantino2018,Aerts2018b,Johnston2019a}. Instead, different 
observables provide constraints on the overall mixing history of a star. 
We will briefly review the types of observations used to constrain 
chemical mixing in the literature.

\subsection{Eclipsing Binaries}
If the two components of a double-lined spectroscopic binary system 
eclipse, one can model the eclipses and radial velocities to determine 
absolute masses and radii of the components. Combined with effective 
temperature estimates from spectroscopy, modelling eclipsing binaries 
provides some of the most precise determination of fundamental stellar 
properties possible, with quoted uncertainties to better than one 
percent on mass and radius \citep{Torres2010,Serenelli2021}. Furthermore, 
binaries afford the assumption that both components have the same age 
and initial chemical composition, providing additional powerful 
constraints. 

The standard approach in the literature is to fit stellar evolution models
(or isochrones) to the observed mass, radius, and temperature of the binary 
components to determine how much internal mixing, if any, is required to explain 
observations of intermediate- to high-mass stars with a convective core 
on the main sequence. This approach has been used to investigate the 
impact of rotational mixing \citep{Brott2011b,Schneider2014,Ekstrom2018}, 
convective penetration \citep{Schroder1997,Pols1997,Ribas2000b,Guinan2000,
Claret2016}, and convective overshooting \citep{Stancliffe2015,Higl2017,
Claret2019,Guglielmo2019,Tkachenko2020}. However, notable works claim 
that chemical mixing efficiency cannot be uniquely determined using this
approach, even with such precise mass and radius estimates \citep{Valle2018,
Constantino2018,Johnston2019b}. Similarly, systems that display apsidal 
motion have been used to investigate chemical mixing through its influence 
on the central condensation of the star over time \citep{Claret2010,Claret2019b}.

We note, however, that the fundamental stellar properties used in this 
modelling method are not directly sensitive to the efficiency of any 
particular mixing mechanism individually. Instead this methodology is 
sensitive to the efficiency of any process which alters the radius and 
temperature of the star. Through the transport of hydrogen from the 
envelope into the core, chemical mixing increases the mass (and luminosity) 
of the core, and extends the main-sequence lifetime. These changes to 
the core translate to changes in the overall temperature and radius of the 
star at a given age, which is what is estimated in binary modelling. 
Furthermore, in addition to the transport of chemicals in the interior, 
very rapid rotation causes changes to the radius and temperature at the 
surface of the star. Thus, modelling the fundamental properties of binaries 
is sensitive to the total internal mixing history which transports fresh 
hydrogen from the radiative region into the convective core throughout 
the stellar lifetime, as well as any modification to the surface properties
due to rapid rotation. To this end, we should be careful to not conflate the 
inferred overall mixing efficiency from binary modelling with the efficiency 
of a single mixing mechanism.

\subsection{Stellar clusters}
Given the assumption that the members of a stellar cluster are formed at 
approximately the same time and with a similar initial chemical composition, 
the observed morphology of the colour-magnitude diagram can be used to 
calibrate internal mixing mechanisms in stellar models. Several studies 
have demonstrated that the width of the main sequence is sensitive to the 
mixing history of stars, and used this sensitivity to estimate the effect 
of overshooting and/or rotational mixing on stellar models \citep{VandenBerg2006,
Castro2014,Martinet2021}. 

Similar to the width of the main sequence, the morphology of the extended 
main-sequence turn off (eMSTO) is used to test the implementation of
mixing mechanisms in stellar models. While the eMSTO phenomenon is commonly 
associated with rapid rotation and its consequences \citep{Dupree2017,Kamann2018,
Bastian2018,Georgy2019}, studies have shown that the eMSTO of young massive 
clusters can be reproduced by stellar evolution models with convective boundary 
mixing calibrated from asteroseismology \citep{Yang2017,Johnston2019c}. 
Alternatively, binary evolution and its byproducts have been invoked as a 
means of explaining the eMSTO phenomenon \citep{Schneider2014b,Gosnell2015,
Beasor2019}. As in the case of eclipsing binaries, this type of analysis 
is not directly sensitive to the efficiency of a given mixing mechanism, 
but rather to the total mixing across the evolution of a star.

\subsection{Asteroseismology}

Regions of the stellar interior can be driven to deviate from 
hydrostatic equilibrium due to a build up of radiation caused 
by partial ionisation zones, the blocking of convective flux, 
or other mechanisms. Depending on where these perturbations
occur within the stellar interior, the restoring force will
be dominated by either the pressure force or the buoyancy
force. If a perturbation is regularly driven, it produces 
a standing pressure or buoyancy wave propagating throughout 
the star, resulting in surface brightness and velocity variations. 
The frequency of the standing wave is determined by the chemical 
and thermodynamical conditions of the regions that the waves 
travel through. Asteroseismology is then the study of the stellar 
interior via the observation of and modelling of stellar oscillations 
\citep{Aerts2010,Aerts2021}.

The frequencies of g~modes (buoyancy waves) in particular are sensitive 
to the near-core chemical gradient \citep{Miglio2008} and rotation rate
\citep{Bouabid2013}. Asteroesismology of stars oscillating in p~modes and
in g~modes with 1.3 $\le$ M $\le$ 24 M$_{\odot}$ has revealed the need 
for a wide range of convective boundary mixing efficiencies to reproduce 
observed pulsation  frequencies \citep{Briquet2007,Handler2013,Moravveji2016,
Schmid2016,Szewczuk2018,Mombarg2019,Pedersen2021}. Additionally, asteroseismology 
of stars pulsating in p~modes (pressure waves) between 1.15-1.5~M$_{\odot}$
has revealed the need for a range of convective mixing efficiencies to 
reproduce observed frequencies \citep{Deheuvels2010,Deheuvels2011,
SilvaAguirre2011,Salmon2012,Angelou2020,Viani2020,Noll2021}. Similar to
binaries and clusters, the modelling of pulsation frequencies (for both 
p and g~modes) is not sensitive to the efficiency of a single mixing 
mechanism. Instead asteroseismology is sensitive to the total influence 
of chemical mixing that alters the core mass and radius, as well as 
the chemical gradient near the core (high order g~modes) and the bulk
density of the star (low order p and g~modes). However, studies have 
demonstrated that g-mode asteroseismology is able to differentiate 
between the morphology and temperature gradient of the near core mixing
profile for models of the same mass evolved to the same core hydrogen 
content \citep{Pedersen2018,Michielsen2019}. 


\section{Convective core masses}
\label{section:core_mass_sample}

Despite the remarkable precision provided by eclipse modelling and/or 
asteroseismology, modelling methodologies cannot uniquely constrain the 
efficiency of an individual mixing mechanism \citep{Valle2018,Aerts2018b}.
Unfortunately, parameter correlations in stellar evolution models 
can produce several models with different masses, ages, and amounts 
of internal mixing that all reproduce observations reasonably well 
within the uncertainties. As such, it is often the case that any 
amount of internal mixing can be used to match observations, even 
when the bulk stellar properties or oscillation frequencies are 
known to a high precision \citep[e.g.,][]{Aerts2018b,Constantino2018,
Johnston2019a, Johnston2019b,Michielsen2019,Sekaran2021}. Furthermore, 
both binary and asteroseismic studies have demonstrated that reasonably 
similar fits to the data can be obtained using models with different 
types of internal mixing mechanisms \citep{Moravveji2015,Claret2017,
Mombarg2019,Pedersen2021} and for different bulk chemical compositions 
\citep{Claret2017,Claret2018}.

Although model degeneracies prevent the direct constraining of mixing 
mechanism efficiencies, \citet{Johnston2019b,Tkachenko2020} demonstrated 
that 1\% precision on mass and radius estimates from eclipsing binaries 
result in inferences on the mass and hydrogen content of the convective 
core to a precision of $\sim10$~\% or better. A similar result has been 
found through asteroseismic analysis of both p and g mode pulsating stars 
with a convective core on the main sequence \citep{Mazumdar2006,Briquet2011,
Johnston2019a,Angelou2020,Mombarg2021,Pedersen2021}.

A few studies have performed comparative model fits between models with an exponentially decaying diffusive overshooting with a radiative temperature gradient in the overshooting region and models with step penetrative convection with an adiabatic expansion of the convective core. These studies demonstrate that the best matching solutions from either model have core masses, masses, and ages, which largely agree within uncertainties \citep{Claret2018,Tkachenko2020,Mombarg2021,Pedersen2021}. However, it should be noted that solutions with different metallicities tend to have different core mass inferences, masses, and ages as well. This highlights that core mass inferences can function as a robust proxy for the overall influence of chemical mixing, even when the efficiency of the implemented mechanism cannot be constrained. Furthermore, this implies that core mass inferences can serve as a robust comparison point for solutions that do not use the same implementation of chemical mixing.

\begin{figure}
   \centering

   \includegraphics[width=0.95\columnwidth]{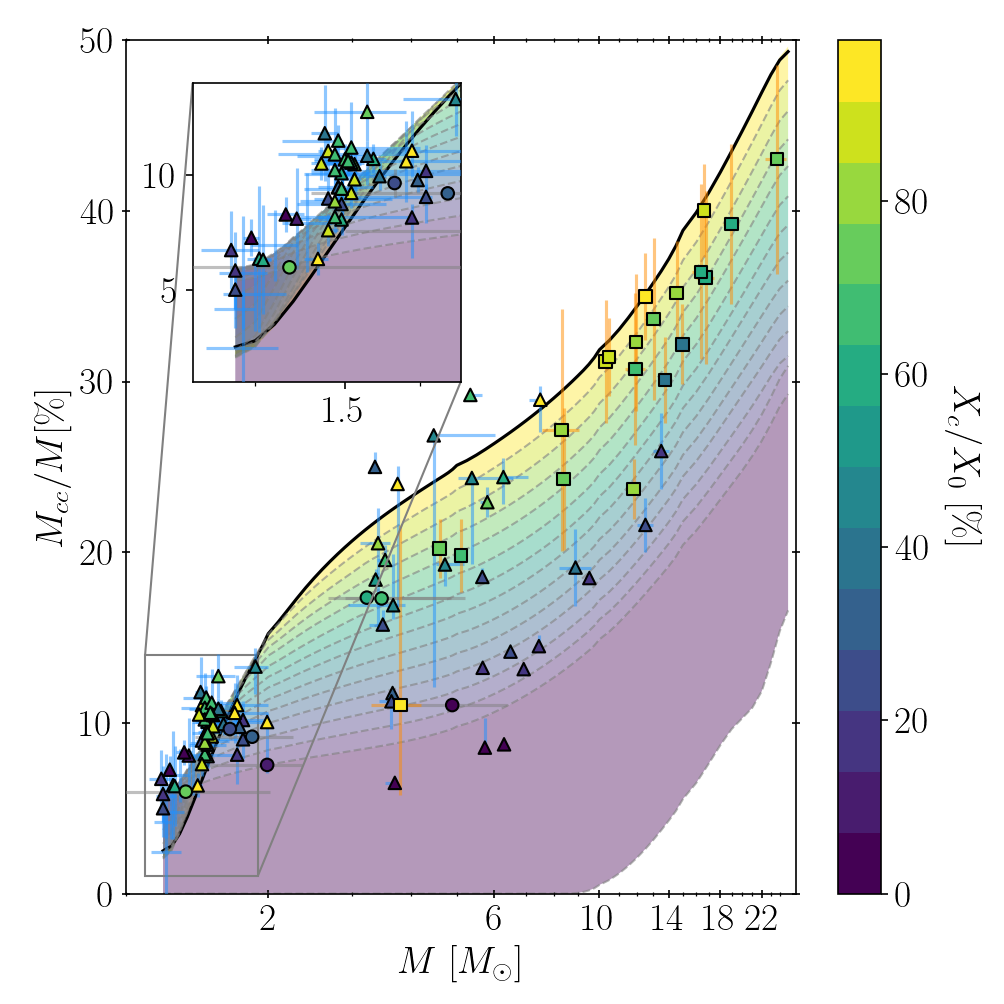}
   \caption{Convective core mass (M$_{cc}$) vs initial mass (M) of the star. Solid and dashed lines 
            correspond to fractional core mass at specified points in evolution. Colour of region 
            corresponds to fractional core hydrogen content as indicated by colour-bar. Colour of 
            individual marker corresponds to inferred fractional core hydrogen content of best 
            model. Squares with orange uncertainties denote values obtained from binary systems, 
            circles with blue uncertainties denote values obtained from binary systems with at least
            one pulsating component, and triangles with grey uncertainties denote values obtained
            from pulsating stars. References for values and their uncertainties are listed in 
            Table~\ref{tab:mccs}.}
              \label{fig:m_vs_mcc_obs}%
\end{figure}

To this end, we collect a large sample of convective core mass and
core hydrogen content inferences (the mass and remaining hydrogen 
content within the Schwarzschild boundary) for core-hydrogen burning 
stars from the literature and show them in Fig.~\ref{fig:m_vs_mcc_obs}. 
The values, their associated uncertainties, and references are 
listed in Table~\ref{tab:mccs}. The core 
mass inferences are plotted against predictions from non-rotating 
models with $Z=0.014$ and $Y=0.276$ \citep[computed with {\sc mesa} 
r10398; ][]{Paxton2019} with a minimum amount of internal chemical 
mixing (included in the form of diffusive exponential overshooting 
with $f_{ov}$=0.005). The models span 1.1\,<\,M\,<\,25~M$_{\odot}$. 
The colour of the background denotes the remaining fraction of the 
core-hydrogen content relative to the amount at the zero-age main sequence. 
The colour of the markers correspond to the inferred fractional 
core-hydrogen content of that star. If the model predictions were 
accurate, the colour of the marker would match the colour of the 
background where the point is situated. Instead, the inferred values 
are systematically shifted upwards with respect to their predicted location, 
indicating that the observations are better reproduced by models with 
more massive convective cores that are less progressed along their 
main-sequence evolution. This sample is composed of inferences from 
p and g mode (and hybrid) pulsating stars, eclipsing binaries, and 
pulsating stars in binaries, spanning a wide mass range 
(1.14\,<\,M\,<\,24~M$_{\odot}$). 

The diversity of this sample {\it i}) confirms the robust need 
for more massive convective cores across a wide stellar mass range, 
and {\it ii}) demonstrates that models including only one 
mixing mechanism with a single efficiency cannot reproduce the 
range of core masses displayed in Fig.~\ref{fig:m_vs_mcc_obs}.
Instead, we find that the range of observations can be reproduced by 
considering stellar models with internal mixing profiles that span a 
wide range of efficiencies. This result is consistent with the fact that 
the observables used in modelling efforts are not directly sensitive to a 
particular chemical mixing mechanism, but rather to the overall influence 
of chemical mixing on the stellar temperature, radius, core properties, 
and near core chemical gradient. Thus, we interpret the diversity
of inferred core masses to result from the combined effect of multiple
mixing mechanisms active in different stars. We adopt a representative 
mixing profile consisting of convective boundary mixing in the form of diffusive 
exponential overshooting \citep{Freytag1996} with $f_{ov}\in[0.005,0.04] H_p$, 
and mixing in the radiative zone induced by internal gravity waves with base 
efficiencies in the range of $D_{\rm IGW}\in [1-100]$~cm$^2$~s$^{-1}$. 
This range is derived by first considering the range of efficiencies
for each mechanism reported in the literature, and then trimming the ranges
to include all values for which we can reproduce the values reported in 
Table~\ref{tab:mccs}. We stress that this representative profile and 
efficiency ranges are only a proxy for the total amount of internal 
chemical mixing, and should not be considered as absolute metric. We 
remark instead that this range of efficiencies can be interpreted as 
arising from different mechanisms producing an overall different effect 
in different stars. 

\section{Implications beyond the main sequence}

The modification of the convective core mass during the main 
sequence propagates throughout the remainder of the star's 
evolution \citep{Stothers1981}. \citet{Kaiser2020} demonstrated 
that variable convective boundary mixing efficiencies lead 
to large differences in the resulting helium-core mass at the
terminal-age main sequence (>30\%), the carbon/oxygen core 
mass at the end of core helium burning (up to 70\%), and in 
the observed surface abundances. These differences necessarily 
have consequences for the stellar end product, although 
\citet{Kaiser2020} only focused on supernova progenitors. Such 
variations in stellar end products have also been demonstrated 
for varying degrees of rapid rotation \citep{Brott2011a,Ekstrom2012}, 
as well as for the inclusion of convective entrainment in 1D 
models \citep{Scott2021}.

\begin{figure}
\centering
\includegraphics[width=0.95\columnwidth]{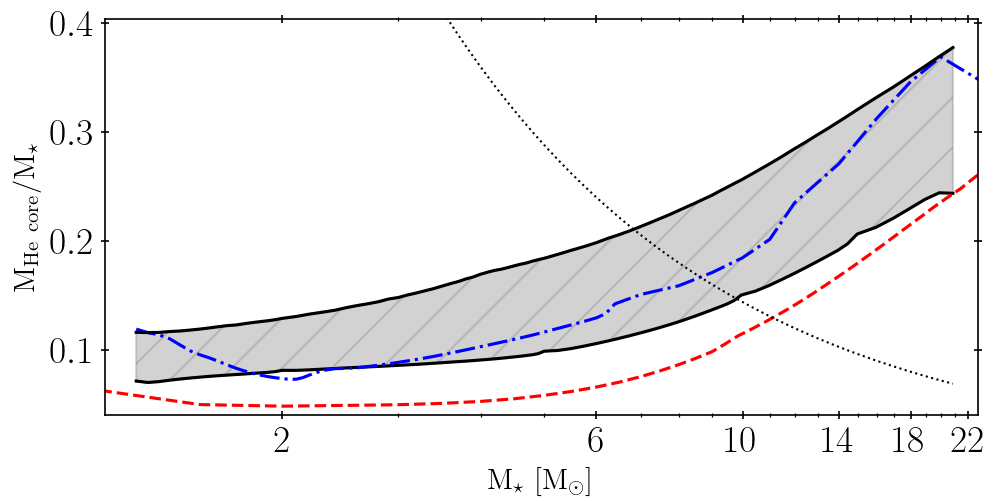}
\caption{Predicted helium core mass at the terminal-age main sequence for 
         models with CBM represented by overshooting and internal gravity 
         wave mixing (grey) compared against predictions from MIST models 
         \citep{Choi2016} in red and PARSEC models \citep{Bressan2012} in 
         blue. The dotted black line denotes the fractional core mass 
         required to produce a remnant helium core with $M=1.44~{\rm M_{\odot}}$.}
          \label{fig:mhe_at_tams}%
\end{figure}

Figure~\ref{fig:mhe_at_tams} shows the predicted values of the 
helium core mass at the end of the main sequence for stars between 
1.1$\le$M$\le$25~M$_{\odot}$. The grey region contains those predicted 
values for models which include the representative range of chemical 
mixing efficiencies which reliably reproduces the range of inferred 
core masses shown in Fig~\ref{fig:m_vs_mcc_obs}. These predictions 
are compared to predictions by non-rotating MIST \citep[red, ][]{Choi2016} 
and PARSEC \citep[blue, ][]{Bressan2012} models. The result of this 
comparison is that we predict a wide range of possible resultant helium 
core masses, compared to the MIST or PARSEC models which adopt a single 
efficiency for one chemical mixing mechanism, and that the predicted 
range of helium core masses increases with mass within the available range.

The prediction of a range of helium core masses has numerous implications 
beyond main-sequence evolution and to other fields of astronomy. Crucially, 
this range suggests that there is not a singular one-to-one relation between 
the birth-mass of the star and the mass of the end-product, as can be seen 
in the range of the helium core masses. \citet{Kaiser2020} demonstrated that 
this can have an impact on the results of supernova simulations, but did not 
consider masses below 15~M$_{\odot}$. The mass limit for a star to produce
a neutron star end-product canonically depends on the remnant core mass exceeding
the Chandrasekhar limit of M$_{\rm crit}\approx1.44$~M$_{\odot}$. Typically, a
remnant core with such a mass is thought to be achieved for stars with initial 
masses M$_{\rm init}>\sim 8-9$~M$_{\odot}$, considering possible accretion onto 
the core at later evolutionary stages \citep{Heger2003,Fryer2005,Camenzind2007,
Kippenhahn2012}. This limiting mass is denoted by the dotted black line in 
Fig.~\ref{fig:mhe_at_tams}. Interestingly we see that this critical remnant 
core mass is already reached for stars with initial masses as low as 
M$_{\rm init}>\sim 7$~M$_{\odot}$ for models with the most internal mixing 
that is considered. This lower limit does not consider possible core mass 
enhancement through later phases of shell burning or convective boundary 
mixing during later core burning phases.

Furthermore, studies have demonstrated that convective boundary mixing is 
favoured to explain observations of evolved single stars \citep{Montalban2013,
Constantino2017,Bossini2017,Arentoft2017,denHartogh2019}, and/or binary 
interaction products, such as subdwarfs \citep{Constantino2018,Ostrowski2021}.
However, the study of binaries and stellar populations is also dependent 
upon considerations of internal mixing in stellar models. Particularly, 
the pathways of binary evolution products depend on the radius and central 
condensation of an evolving star to determine the probability, duration, 
and efficiency of mass transfer.

%

\section{Conclusions}
\label{section:conclusions}

It is easy to review the results of all the attempts to calibrate 
different internal mixing mechanisms and think that they disagree. 
However, given the nature of the sensitivity of modern observables
combined with the limitations of 1D stellar models, we argue that 
these results do agree. Irrespective of the type of internal mixing
mechanism implemented, models require a mechanism to modify the 
core mass, internal chemical gradients, and surface properties in 
order to reproduce observations. This is supported by the fact that 
different studies have been able to independently reproduce the behaviour 
of commonly used observables with different implementations of mixing
mechanisms such as convective penetration, overshooting, entrainment, 
rotation, and internal gravity waves \citep{Brott2011b,Ekstrom2012,
Staritsin2013,Moravveji2015,Claret2017,Tkachenko2020,Pedersen2021,
Scott2021}.

Consulting the diversity of results from the literature, it is clear 
that a wide range of internal mixing efficiencies are required to 
reproduce observations even when considering a single mixing mechanism. 
When considering a single efficiency for a single mixing mechanism, 
models predictions are unintentionally biased. Based on the diverse 
sample of inferred core masses we have collected, we instead argue 
that a range of efficiencies needs to be considered to function as a
proxy for the various mixing processes present at the convective
boundary which are commonly ignored in standard 1D models. As this 
in turn produces a range of helium core masses at the terminal-age 
main sequence, this range of efficiencies should be accounted for 
in evolutionary calculations for both single and binary stars, 
nucleosynthetic yield predictions, and population synthesis efforts 
beyond the main sequence.

\begin{acknowledgements}
CJ would like to thank the referee for their comments which improved the 
manuscript. CJ thanks M.G. Pedersen, J.S.G. Mombarg, G. Angelou, L. Viani, 
and C. Neiner for sharing their results to be included in this manuscript, as
well as D.M. Bowman, A. Tkachenko, and C. Aerts for their useful discussion
on topics concerning asteroseismology and chemical mixing profiles, and M. 
Abdul-Masih and A. Escorza for their comments on the manuscript. This 
work has received funding from NOVA, the European Research Council under the 
European Union's Horizon 2020 research and innovation programme (N$^\circ$670519:MAMSIE), 
and from the Research Foundation Flanders under grant agreement G0A2917N (BlackGEM).
The computational resources and services used in this work were provided by 
the VSC (Flemish Supercomputer Center), funded by the Research Foundation - 
Flanders (FWO) and the Flemish Government - department EWI to PI Johnston.
\end{acknowledgements}

\bibliographystyle{aa}
\bibliography{astroBib.bib}

\begin{thebibliography}{99}
\expandafter\ifx\csname natexlab\endcsname\relax\def\natexlab#1{#1}\fi

\bibitem[{{Aerts}(2021)}]{Aerts2021}
{Aerts}, C. 2021, Reviews of Modern Physics, 93, 015001

\bibitem[{{Aerts} {et~al.}(2010){Aerts}, {Christensen-Dalsgaard}, \&
  {Kurtz}}]{Aerts2010}
{Aerts}, C., {Christensen-Dalsgaard}, J., \& {Kurtz}, D.~W. 2010,
  {Asteroseismology, Astronomy and Astrophysics Library} (Springer-Verlag,
  Heidelberg)

\bibitem[{{Aerts} {et~al.}(2018){Aerts}, {Molenberghs}, {Michielsen},
  {Pedersen}, {Bj{\"o}rklund}, {Johnston}, {Mombarg}, {Bowman}, {Buysschaert},
  {P{\'a}pics}, {Sekaran}, {Sundqvist}, {Tkachenko}, {Truyaert}, {Van Reeth},
  \& {Vermeyen}}]{Aerts2018b}
{Aerts}, C., {Molenberghs}, G., {Michielsen}, M., {et~al.} 2018, \apjs, 237, 15

\bibitem[{{Angelou} {et~al.}(2020){Angelou}, {Bellinger}, {Hekker}, {Mints},
  {Elsworth}, {Basu}, \& {Weiss}}]{Angelou2020}
{Angelou}, G.~C., {Bellinger}, E.~P., {Hekker}, S., {et~al.} 2020, \mnras, 493,
  4987

\bibitem[{{Arentoft} {et~al.}(2017){Arentoft}, {Brogaard}, {Jessen-Hansen},
  {Silva Aguirre}, {Kjeldsen}, {Mosumgaard}, \& {Sandquist}}]{Arentoft2017}
{Arentoft}, T., {Brogaard}, K., {Jessen-Hansen}, J., {et~al.} 2017, \apj, 838,
  115

\bibitem[{{Arnett} {et~al.}(2019){Arnett}, {Meakin}, {Hirschi}, {Cristini},
  {Georgy}, {Campbell}, {Scott}, {Kaiser}, {Viallet}, \&
  {Moc{\'a}k}}]{Arnett2019}
{Arnett}, W.~D., {Meakin}, C., {Hirschi}, R., {et~al.} 2019, \apj, 882, 18

\bibitem[{{Arnett} {et~al.}(2015){Arnett}, {Meakin}, {Viallet}, {Campbell},
  {Lattanzio}, \& {Moc{\'a}k}}]{Arnett2015}
{Arnett}, W.~D., {Meakin}, C., {Viallet}, M., {et~al.} 2015, \apj, 809, 30

\bibitem[{{Bastian} {et~al.}(2018){Bastian}, {Kamann}, {Cabrera-Ziri},
  {Georgy}, {Ekstr{\"o}m}, {Charbonnel}, {de Juan Ovelar}, \&
  {Usher}}]{Bastian2018}
{Bastian}, N., {Kamann}, S., {Cabrera-Ziri}, I., {et~al.} 2018, \mnras, 480,
  3739

\bibitem[{{Beasor} {et~al.}(2019){Beasor}, {Davies}, {Smith}, \&
  {Bastian}}]{Beasor2019}
{Beasor}, E.~R., {Davies}, B., {Smith}, N., \& {Bastian}, N. 2019, \mnras, 486,
  266

\bibitem[{{Bertelli} {et~al.}(1985){Bertelli}, {Bressan}, \&
  {Chiosi}}]{Bertelli1985}
{Bertelli}, G., {Bressan}, A.~G., \& {Chiosi}, C. 1985, \aap, 150, 33

\bibitem[{{B{\"o}hm-Vitense}(1958)}]{BohmVitense1958}
{B{\"o}hm-Vitense}, E. 1958, \zap, 46, 108

\bibitem[{{Bossini} {et~al.}(2017){Bossini}, {Miglio}, {Salaris}, {Vrard},
  {Cassisi}, {Mosser}, {Montalb{\'a}n}, {Girardi}, {Noels}, {Bressan},
  {Pietrinferni}, \& {Tayar}}]{Bossini2017}
{Bossini}, D., {Miglio}, A., {Salaris}, M., {et~al.} 2017, \mnras, 469, 4718

\bibitem[{{Bouabid} {et~al.}(2013){Bouabid}, {Dupret}, {Salmon},
  {Montalb{\'a}n}, {Miglio}, \& {Noels}}]{Bouabid2013}
{Bouabid}, M.~P., {Dupret}, M.~A., {Salmon}, S., {et~al.} 2013, \mnras, 429,
  2500

\bibitem[{{Bressan} {et~al.}(2012){Bressan}, {Marigo}, {Girardi}, {Salasnich},
  {Dal Cero}, {Rubele}, \& {Nanni}}]{Bressan2012}
{Bressan}, A., {Marigo}, P., {Girardi}, L., {et~al.} 2012, \mnras, 427, 127

\bibitem[{{Bressan} {et~al.}(1981){Bressan}, {Chiosi}, \&
  {Bertelli}}]{Bressan1981}
{Bressan}, A.~G., {Chiosi}, C., \& {Bertelli}, G. 1981, \aap, 102, 25

\bibitem[{{Briquet} {et~al.}(2011){Briquet}, {Aerts}, {Baglin}, {Nieva},
  {Degroote}, {Przybilla}, {Noels}, {Schiller}, {Vu{\v{c}}kovi{\'c}}, {Oreiro},
  {Smolders}, {Auvergne}, {Baudin}, {Catala}, {Michel}, \&
  {Samadi}}]{Briquet2011}
{Briquet}, M., {Aerts}, C., {Baglin}, A., {et~al.} 2011, \aap, 527, A112

\bibitem[{{Briquet} {et~al.}(2007){Briquet}, {Morel}, {Thoul}, {Scuflaire},
  {Miglio}, {Montalb{\'a}n}, {Dupret}, \& {Aerts}}]{Briquet2007}
{Briquet}, M., {Morel}, T., {Thoul}, A., {et~al.} 2007, \mnras, 381, 1482

\bibitem[{{Brott} {et~al.}(2011{\natexlab{a}}){Brott}, {de Mink}, {Cantiello},
  {Langer}, {de Koter}, {Evans}, {Hunter}, {Trundle}, \& {Vink}}]{Brott2011a}
{Brott}, I., {de Mink}, S.~E., {Cantiello}, M., {et~al.} 2011{\natexlab{a}},
  \aap, 530, A115

\bibitem[{{Brott} {et~al.}(2011{\natexlab{b}}){Brott}, {Evans}, {Hunter}, {de
  Koter}, {Langer}, {Dufton}, {Cantiello}, {Trundle}, {Lennon}, {de Mink},
  {Yoon}, \& {Anders}}]{Brott2011b}
{Brott}, I., {Evans}, C.~J., {Hunter}, I., {et~al.} 2011{\natexlab{b}}, \aap,
  530, A116

\bibitem[{{Buysschaert} {et~al.}(2018){Buysschaert}, {Aerts}, {Bowman},
  {Johnston}, {Van Reeth}, {Pedersen}, {Mathis}, \&
  {Neiner}}]{Buysschaert2018b}
{Buysschaert}, B., {Aerts}, C., {Bowman}, D.~M., {et~al.} 2018, \aap, 616, A148

\bibitem[{{Camenzind}(2007)}]{Camenzind2007}
{Camenzind}, M. 2007, {Compact objects in astrophysics : white dwarfs, neutron
  stars, and black holes}

\bibitem[{{Castro} {et~al.}(2014){Castro}, {Fossati}, {Langer},
  {Sim{\'o}n-D{\'\i}az}, {Schneider}, \& {Izzard}}]{Castro2014}
{Castro}, N., {Fossati}, L., {Langer}, N., {et~al.} 2014, \aap, 570, L13

\bibitem[{{Choi} {et~al.}(2016){Choi}, {Dotter}, {Conroy}, {Cantiello},
  {Paxton}, \& {Johnson}}]{Choi2016}
{Choi}, J., {Dotter}, A., {Conroy}, C., {et~al.} 2016, \apj, 823, 102

\bibitem[{{Claret}(2019)}]{Claret2019b}
{Claret}, A. 2019, \aap, 628, A29

\bibitem[{{Claret} \& {Gim{\'e}nez}(2010)}]{Claret2010}
{Claret}, A. \& {Gim{\'e}nez}, A. 2010, \aap, 519, A57

\bibitem[{{Claret} \& {Torres}(2016)}]{Claret2016}
{Claret}, A. \& {Torres}, G. 2016, \aap, 592, A15

\bibitem[{{Claret} \& {Torres}(2017)}]{Claret2017}
{Claret}, A. \& {Torres}, G. 2017, \apj, 849, 18

\bibitem[{{Claret} \& {Torres}(2018)}]{Claret2018}
{Claret}, A. \& {Torres}, G. 2018, \apj, 859, 100

\bibitem[{{Claret} \& {Torres}(2019)}]{Claret2019}
{Claret}, A. \& {Torres}, G. 2019, \apj, 876, 134

\bibitem[{{Constantino} \& {Baraffe}(2018)}]{Constantino2018}
{Constantino}, T. \& {Baraffe}, I. 2018, \aap, 618, A177

\bibitem[{{Constantino} {et~al.}(2017){Constantino}, {Campbell}, \&
  {Lattanzio}}]{Constantino2017}
{Constantino}, T., {Campbell}, S.~W., \& {Lattanzio}, J.~C. 2017, \mnras, 472,
  4900

\bibitem[{{Costa} {et~al.}(2019){Costa}, {Girardi}, {Bressan}, {Marigo},
  {Rodrigues}, {Chen}, {Lanza}, \& {Goudfrooij}}]{Guglielmo2019}
{Costa}, G., {Girardi}, L., {Bressan}, A., {et~al.} 2019, \mnras, 485, 4641

\bibitem[{{Cristini} {et~al.}(2017){Cristini}, {Meakin}, {Hirschi}, {Arnett},
  {Georgy}, {Viallet}, \& {Walkington}}]{Cristini2017}
{Cristini}, A., {Meakin}, C., {Hirschi}, R., {et~al.} 2017, \mnras, 471, 279

\bibitem[{{Deheuvels} {et~al.}(2010){Deheuvels}, {Bruntt}, {Michel}, {Barban},
  {Verner}, {R{\'e}gulo}, {Mosser}, {Mathur}, {Gaulme}, {Garcia}, {Boumier},
  {Appourchaux}, {Samadi}, {Catala}, {Baudin}, {Baglin}, {Auvergne},
  {Roxburgh}, \& {P{\'e}rez Hern{\'a}ndez}}]{Deheuvels2010}
{Deheuvels}, S., {Bruntt}, H., {Michel}, E., {et~al.} 2010, \aap, 515, A87

\bibitem[{{Deheuvels} \& {Michel}(2011)}]{Deheuvels2011}
{Deheuvels}, S. \& {Michel}, E. 2011, \aap, 535, A91

\bibitem[{{den Hartogh} {et~al.}(2019){den Hartogh}, {Eggenberger}, \&
  {Hirschi}}]{denHartogh2019}
{den Hartogh}, J.~W., {Eggenberger}, P., \& {Hirschi}, R. 2019, \aap, 622, A187

\bibitem[{{Dupree} {et~al.}(2017){Dupree}, {Dotter}, {Johnson}, {Marino},
  {Milone}, {Bailey}, {Crane}, {Mateo}, \& {Olszewski}}]{Dupree2017}
{Dupree}, A.~K., {Dotter}, A., {Johnson}, C.~I., {et~al.} 2017, \apjl, 846, L1

\bibitem[{{Ekstr{\"o}m} {et~al.}(2012){Ekstr{\"o}m}, {Georgy}, {Eggenberger},
  {Meynet}, {Mowlavi}, {Wyttenbach}, {Granada}, {Decressin}, {Hirschi},
  {Frischknecht}, {Charbonnel}, \& {Maeder}}]{Ekstrom2012}
{Ekstr{\"o}m}, S., {Georgy}, C., {Eggenberger}, P., {et~al.} 2012, \aap, 537,
  A146

\bibitem[{{Ekstr{\"o}m} {et~al.}(2018){Ekstr{\"o}m}, {Meynet}, {Georgy}, \&
  {Granada}}]{Ekstrom2018}
{Ekstr{\"o}m}, S., {Meynet}, G., {Georgy}, C., \& {Granada}, A. 2018, Memorie
  della Societa Astronomica Italiana, 89, 50

\bibitem[{{Freytag} {et~al.}(1996){Freytag}, {Ludwig}, \&
  {Steffen}}]{Freytag1996}
{Freytag}, B., {Ludwig}, H.~G., \& {Steffen}, M. 1996, \aap, 313, 497

\bibitem[{{Fryer} \& {Hungerford}(2005)}]{Fryer2005}
{Fryer}, C.~L. \& {Hungerford}, A. 2005, in NATO Advanced Study Institute (ASI)
  Series B, Vol. 210, The Electromagnetic Spectrum of Neutron Stars, 3

\bibitem[{{Georgy} {et~al.}(2019){Georgy}, {Charbonnel}, {Amard}, {Bastian},
  {Ekstr{\"o}m}, {Lardo}, {Palacios}, {Eggenberger}, {Cabrera-Ziri}, {Gallet},
  \& {Lagarde}}]{Georgy2019}
{Georgy}, C., {Charbonnel}, C., {Amard}, L., {et~al.} 2019, \aap, 622, A66

\bibitem[{{Gosnell} {et~al.}(2015){Gosnell}, {Mathieu}, {Geller}, {Sills},
  {Leigh}, \& {Knigge}}]{Gosnell2015}
{Gosnell}, N.~M., {Mathieu}, R.~D., {Geller}, A.~M., {et~al.} 2015, \apj, 814,
  163

\bibitem[{{Guinan} {et~al.}(2000){Guinan}, {Ribas}, {Fitzpatrick},
  {Gim{\'e}nez}, {Jordi}, {McCook}, \& {Popper}}]{Guinan2000}
{Guinan}, E.~F., {Ribas}, I., {Fitzpatrick}, E.~L., {et~al.} 2000, \apj, 544,
  409

\bibitem[{{Handler}(2013)}]{Handler2013}
{Handler}, G. 2013, {Asteroseismology}, ed. T.~D. {Oswalt} \& M.~A. {Barstow},
  Vol.~4, 207

\bibitem[{{Heger} {et~al.}(2003){Heger}, {Fryer}, {Woosley}, {Langer}, \&
  {Hartmann}}]{Heger2003}
{Heger}, A., {Fryer}, C.~L., {Woosley}, S.~E., {Langer}, N., \& {Hartmann},
  D.~H. 2003, \apj, 591, 288

\bibitem[{{Higl} {et~al.}(2021){Higl}, {M{\"u}ller}, \& {Weiss}}]{Higl2021}
{Higl}, J., {M{\"u}ller}, E., \& {Weiss}, A. 2021, \aap, 646, A133

\bibitem[{{Higl} \& {Weiss}(2017)}]{Higl2017}
{Higl}, J. \& {Weiss}, A. 2017, \aap, 608, A62

\bibitem[{{Hirschi} {et~al.}(2014){Hirschi}, {den Hartogh}, {Cristini},
  {Georgy}, \& {Pignatari}}]{Hirschi2014}
{Hirschi}, R., {den Hartogh}, J., {Cristini}, A., {Georgy}, C., \& {Pignatari},
  M. 2014, in XIII Nuclei in the Cosmos (NIC XIII), 1

\bibitem[{{Johnston} {et~al.}(2019{\natexlab{a}}){Johnston}, {Aerts},
  {Pedersen}, \& {Bastian}}]{Johnston2019c}
{Johnston}, C., {Aerts}, C., {Pedersen}, M.~G., \& {Bastian}, N.
  2019{\natexlab{a}}, \aap, 632, A74

\bibitem[{{Johnston} {et~al.}(2021){Johnston}, {Aimar}, {Abdul-Masih},
  {Bowman}, {White}, {Hawcroft}, {Sana}, {Sekaran}, {Dsilva}, {Tkachenko}, \&
  {Aerts}}]{Johnston2021}
{Johnston}, C., {Aimar}, N., {Abdul-Masih}, M., {et~al.} 2021, \mnras, 503,
  1124

\bibitem[{{Johnston} {et~al.}(2019{\natexlab{b}}){Johnston}, {Pavlovski}, \&
  {Tkachenko}}]{Johnston2019b}
{Johnston}, C., {Pavlovski}, K., \& {Tkachenko}, A. 2019{\natexlab{b}}, \aap,
  628, A25

\bibitem[{{Johnston} {et~al.}(2019{\natexlab{c}}){Johnston}, {Tkachenko},
  {Aerts}, {Molenberghs}, {Bowman}, {Pedersen}, {Buysschaert}, \&
  {P{\'a}pics}}]{Johnston2019a}
{Johnston}, C., {Tkachenko}, A., {Aerts}, C., {et~al.} 2019{\natexlab{c}},
  \mnras, 482, 1231

\bibitem[{{Kaiser} {et~al.}(2020){Kaiser}, {Hirschi}, {Arnett}, {Georgy},
  {Scott}, \& {Cristini}}]{Kaiser2020}
{Kaiser}, E.~A., {Hirschi}, R., {Arnett}, W.~D., {et~al.} 2020, \mnras, 496,
  1967

\bibitem[{{Kamann} {et~al.}(2018){Kamann}, {Bastian}, {Husser}, {Martocchia},
  {Usher}, {den Brok}, {Dreizler}, {Kelz}, {Krajnovi{\'c}}, {Richard},
  {Steinmetz}, \& {Weilbacher}}]{Kamann2018}
{Kamann}, S., {Bastian}, N., {Husser}, T.~O., {et~al.} 2018, \mnras, 480, 1689

\bibitem[{{Kippenhahn} {et~al.}(2012){Kippenhahn}, {Weigert}, \&
  {Weiss}}]{Kippenhahn2012}
{Kippenhahn}, R., {Weigert}, A., \& {Weiss}, A. 2012, {Stellar Structure and
  Evolution} (Springer-Verlag Berlin Heidelberg)

\bibitem[{{Langer}(2012)}]{Langer2012}
{Langer}, N. 2012, \araa, 50, 107

\bibitem[{{Maeder}(2009)}]{Maeder2009}
{Maeder}, A. 2009, {Physics, Formation and Evolution of Rotating Stars}
  (Springer-Verlag Berlin Heidelberg)

\bibitem[{{Maeder} \& {Meynet}(1987)}]{Maeder1987}
{Maeder}, A. \& {Meynet}, G. 1987, \aap, 182, 243

\bibitem[{{Martinet} {et~al.}(2021){Martinet}, {Meynet}, {Ekstr{\"o}m},
  {Sim{\'o}n-D{\'\i}az}, {Holgado}, {Castro}, {Georgy}, {Eggenberger},
  {Buldgen}, {Salmon}, {Hirschi}, {Groh}, {Farrell}, \&
  {Murphy}}]{Martinet2021}
{Martinet}, S., {Meynet}, G., {Ekstr{\"o}m}, S., {et~al.} 2021, \aap, 648, A126

\bibitem[{{Mazumdar} {et~al.}(2006){Mazumdar}, {Briquet}, {Desmet}, \&
  {Aerts}}]{Mazumdar2006}
{Mazumdar}, A., {Briquet}, M., {Desmet}, M., \& {Aerts}, C. 2006, \aap, 459,
  589

\bibitem[{{Meakin} \& {Arnett}(2007)}]{Meakin2007}
{Meakin}, C.~A. \& {Arnett}, D. 2007, \apj, 667, 448

\bibitem[{{Michielsen} {et~al.}(2019){Michielsen}, {Pedersen}, {Augustson},
  {Mathis}, \& {Aerts}}]{Michielsen2019}
{Michielsen}, M., {Pedersen}, M.~G., {Augustson}, K.~C., {Mathis}, S., \&
  {Aerts}, C. 2019, \aap, 628, A76

\bibitem[{{Miglio} {et~al.}(2008){Miglio}, {Montalb{\'a}n}, {Noels}, \&
  {Eggenberger}}]{Miglio2008}
{Miglio}, A., {Montalb{\'a}n}, J., {Noels}, A., \& {Eggenberger}, P. 2008,
  \mnras, 386, 1487

\bibitem[{{Mombarg} {et~al.}(2021){Mombarg}, {Van Reeth}, \&
  {Aerts}}]{Mombarg2021}
{Mombarg}, J.~S.~G., {Van Reeth}, T., \& {Aerts}, C. 2021, \aap, 650, A58

\bibitem[{{Mombarg} {et~al.}(2019){Mombarg}, {Van Reeth}, {Pedersen},
  {Molenberghs}, {Bowman}, {Johnston}, {Tkachenko}, \& {Aerts}}]{Mombarg2019}
{Mombarg}, J.~S.~G., {Van Reeth}, T., {Pedersen}, M.~G., {et~al.} 2019, \mnras,
  485, 3248

\bibitem[{{Montalb{\'a}n} {et~al.}(2013){Montalb{\'a}n}, {Miglio}, {Noels},
  {Dupret}, {Scuflaire}, \& {Ventura}}]{Montalban2013}
{Montalb{\'a}n}, J., {Miglio}, A., {Noels}, A., {et~al.} 2013, \apj, 766, 118

\bibitem[{{Moravveji} {et~al.}(2015){Moravveji}, {Aerts}, {P{\'a}pics},
  {Triana}, \& {Vandoren}}]{Moravveji2015}
{Moravveji}, E., {Aerts}, C., {P{\'a}pics}, P.~I., {Triana}, S.~A., \&
  {Vandoren}, B. 2015, \aap, 580, A27

\bibitem[{{Moravveji} {et~al.}(2016){Moravveji}, {Townsend}, {Aerts}, \&
  {Mathis}}]{Moravveji2016}
{Moravveji}, E., {Townsend}, R.~H.~D., {Aerts}, C., \& {Mathis}, S. 2016, \apj,
  823, 130

\bibitem[{{Noll} {et~al.}(2021){Noll}, {Deheuvels}, \& {Ballot}}]{Noll2021}
{Noll}, A., {Deheuvels}, S., \& {Ballot}, J. 2021, \aap, 647, A187

\bibitem[{{Ostrowski} {et~al.}(2021){Ostrowski}, {Baran}, {Sanjayan}, \&
  {Sahoo}}]{Ostrowski2021}
{Ostrowski}, J., {Baran}, A.~S., {Sanjayan}, S., \& {Sahoo}, S.~K. 2021,
  \mnras, 503, 4646

\bibitem[{{Paxton} {et~al.}(2019){Paxton}, {Smolec}, {Schwab}, {Gautschy},
  {Bildsten}, {Cantiello}, {Dotter}, {Farmer}, {Goldberg}, {Jermyn}, {Kanbur},
  {Marchant}, {Thoul}, {Townsend}, {Wolf}, {Zhang}, \& {Timmes}}]{Paxton2019}
{Paxton}, B., {Smolec}, R., {Schwab}, J., {et~al.} 2019, \apjs, 243, 10

\bibitem[{{Pedersen} {et~al.}(2021){Pedersen}, {Aerts}, {P{\'a}pics},
  {Michielsen}, {Gebruers}, {Rogers}, {Molenberghs}, {Burssens}, {Garcia}, \&
  {Bowman}}]{Pedersen2021}
{Pedersen}, M.~G., {Aerts}, C., {P{\'a}pics}, P.~I., {et~al.} 2021, Nature
  Astronomy [\eprint[arXiv]{2105.04533}]

\bibitem[{{Pedersen} {et~al.}(2018){Pedersen}, {Aerts}, {P{\'a}pics}, \&
  {Rogers}}]{Pedersen2018}
{Pedersen}, M.~G., {Aerts}, C., {P{\'a}pics}, P.~I., \& {Rogers}, T.~M. 2018,
  \aap, 614, A128

\bibitem[{{Pols} {et~al.}(1997){Pols}, {Tout}, {Schroder}, \&
  {Eggleton}}]{Pols1997}
{Pols}, O.~R., {Tout}, C.~A., {Schroder}, K.-P., \& {Eggleton}. 1997, \mnras,
  289, 869

\bibitem[{{Renzini}(1987)}]{Renzini1987}
{Renzini}, A. 1987, \aap, 188, 49

\bibitem[{{Ribas} {et~al.}(2000){Ribas}, {Jordi}, \&
  {Gim{\'e}nez}}]{Ribas2000b}
{Ribas}, I., {Jordi}, C., \& {Gim{\'e}nez}, {\'A}. 2000, \mnras, 318, L55

\bibitem[{{Rogers} \& {McElwaine}(2017)}]{Rogers2017}
{Rogers}, T.~M. \& {McElwaine}, J.~N. 2017, \apjl, 848, L1

\bibitem[{{Salaris} \& {Cassisi}(2017)}]{Salaris2017}
{Salaris}, M. \& {Cassisi}, S. 2017, Royal Society Open Science, 4, 170192

\bibitem[{{Salmon} {et~al.}(2012){Salmon}, {Montalb{\'a}n}, {Morel}, {Miglio},
  {Dupret}, \& {Noels}}]{Salmon2012}
{Salmon}, S., {Montalb{\'a}n}, J., {Morel}, T., {et~al.} 2012, \mnras, 422,
  3460

\bibitem[{{Schmid} \& {Aerts}(2016)}]{Schmid2016}
{Schmid}, V.~S. \& {Aerts}, C. 2016, \aap, 592, A116

\bibitem[{{Schneider} {et~al.}(2014{\natexlab{a}}){Schneider}, {Izzard}, {de
  Mink}, {Langer}, {Stolte}, {de Koter}, {Gvaramadze}, {Hu{\ss}mann},
  {Liermann}, \& {Sana}}]{Schneider2014b}
{Schneider}, F.~R.~N., {Izzard}, R.~G., {de Mink}, S.~E., {et~al.}
  2014{\natexlab{a}}, \apj, 780, 117

\bibitem[{{Schneider} {et~al.}(2014{\natexlab{b}}){Schneider}, {Langer}, {de
  Koter}, {Brott}, {Izzard}, \& {Lau}}]{Schneider2014}
{Schneider}, F.~R.~N., {Langer}, N., {de Koter}, A., {et~al.}
  2014{\natexlab{b}}, \aap, 570, A66

\bibitem[{{Schroder} {et~al.}(1997){Schroder}, {Pols}, \&
  {Eggleton}}]{Schroder1997}
{Schroder}, K.-P., {Pols}, O.~R., \& {Eggleton}, P.~P. 1997, \mnras, 285, 696

\bibitem[{{Scott} {et~al.}(2021){Scott}, {Hirschi}, {Georgy}, {Arnett},
  {Meakin}, {Kaiser}, {Ekstr{\"o}m}, \& {Yusof}}]{Scott2021}
{Scott}, L.~J.~A., {Hirschi}, R., {Georgy}, C., {et~al.} 2021, \mnras, 503,
  4208

\bibitem[{{Sekaran} {et~al.}(2021){Sekaran}, {Tkachenko}, {Johnston}, \&
  {Aerts}}]{Sekaran2021}
{Sekaran}, S., {Tkachenko}, A., {Johnston}, C., \& {Aerts}, C. 2021, \aap, 648,
  A91

\bibitem[{{Serenelli} {et~al.}(2021){Serenelli}, {Weiss}, {Aerts}, {Angelou},
  {Baroch}, {Bastian}, {Beck}, {Bergemann}, {Bestenlehner}, {Czekala},
  {Elias-Rosa}, {Escorza}, {Van Eylen}, {Feuillet}, {Gandolfi}, {Gieles},
  {Girardi}, {Lebreton}, {Lodieu}, {Martig}, {Miller Bertolami}, {Mombarg},
  {Morales}, {Moya}, {Nsamba}, {Pavlovski}, {Pedersen}, {Ribas}, {Schneider},
  {Silva Aguirre}, {Stassun}, {Tolstoy}, {Tremblay}, \&
  {Zwintz}}]{Serenelli2021}
{Serenelli}, A., {Weiss}, A., {Aerts}, C., {et~al.} 2021, \aapr, 29, 4

\bibitem[{{Silva Aguirre} {et~al.}(2011){Silva Aguirre}, {Ballot}, {Serenelli},
  \& {Weiss}}]{SilvaAguirre2011}
{Silva Aguirre}, V., {Ballot}, J., {Serenelli}, A.~M., \& {Weiss}, A. 2011,
  \aap, 529, A63

\bibitem[{{Stancliffe} {et~al.}(2015){Stancliffe}, {Fossati}, {Passy}, \&
  {Schneider}}]{Stancliffe2015}
{Stancliffe}, R.~J., {Fossati}, L., {Passy}, J.~C., \& {Schneider}, F.~R.~N.
  2015, \aap, 575, A117

\bibitem[{{Staritsin}(2013)}]{Staritsin2013}
{Staritsin}, E.~I. 2013, Astronomy Reports, 57, 380

\bibitem[{{Stothers} \& {Chin}(1981)}]{Stothers1981}
{Stothers}, R. \& {Chin}, C.~W. 1981, \apj, 247, 1063

\bibitem[{{Szewczuk} \& {Daszy{\'n}ska-Daszkiewicz}(2018)}]{Szewczuk2018}
{Szewczuk}, W. \& {Daszy{\'n}ska-Daszkiewicz}, J. 2018, \mnras, 478, 2243

\bibitem[{{Tkachenko} {et~al.}(2020){Tkachenko}, {Pavlovski}, {Johnston},
  {Pedersen}, {Michielsen}, {Bowman}, {Southworth}, {Tsymbal}, \&
  {Aerts}}]{Tkachenko2020}
{Tkachenko}, A., {Pavlovski}, K., {Johnston}, C., {et~al.} 2020, \aap, 637, A60

\bibitem[{{Torres} {et~al.}(2010){Torres}, {Andersen}, \&
  {Gim{\'e}nez}}]{Torres2010}
{Torres}, G., {Andersen}, J., \& {Gim{\'e}nez}, A. 2010, ARA\&A, 18, 67

\bibitem[{{Valle} {et~al.}(2017){Valle}, {Dell'Omodarme}, {Prada Moroni}, \&
  {Degl'Innocenti}}]{Valle2017}
{Valle}, G., {Dell'Omodarme}, M., {Prada Moroni}, P.~G., \& {Degl'Innocenti},
  S. 2017, \aap, 600, A41

\bibitem[{{Valle} {et~al.}(2018){Valle}, {Dell'Omodarme}, {Prada Moroni}, \&
  {Degl'Innocenti}}]{Valle2018}
{Valle}, G., {Dell'Omodarme}, M., {Prada Moroni}, P.~G., \& {Degl'Innocenti},
  S. 2018, \aap, 615, A62

\bibitem[{{VandenBerg} {et~al.}(2006){VandenBerg}, {Bergbusch}, \&
  {Dowler}}]{VandenBerg2006}
{VandenBerg}, D.~A., {Bergbusch}, P.~A., \& {Dowler}, P.~D. 2006, \apjs, 162,
  375

\bibitem[{{Viani} \& {Basu}(2020)}]{Viani2020}
{Viani}, L.~S. \& {Basu}, S. 2020, \apj, 904, 22

\bibitem[{{Yang} \& {Tian}(2017)}]{Yang2017}
{Yang}, W. \& {Tian}, Z. 2017, \apj, 836, 102

\end{thebibliography}

\begin{appendix}
\section{Sample}

\begin{table*}
\small
\caption{Sample of estimated stellar masses and inferred convective core masses and fractional core-hydrogen content.\label{tab:mccs}}
\centering
\begin{tabular}{lcccc|lcccc}
\hline\hline
Star&Mass~[M$_{\odot}$]&M$_{\rm cc}$~[M$_{\odot}$]&X$_c$/X$_{ini}$ & & Star&Mass~[M$_{\odot}$]&M$_{\rm cc}$~[M$_{\odot}$]&X$_c$/X$_{ini}$ & \\
\hline
KIC 12009504 & $1.19\pm0.07$                   & $0.08 \pm0.02$                  & 0.21 & (a)  & KIC 7760680     & $3.37\substack{+0.01 \\ -0.06}$ & $0.62\substack{+0.02 \\ -0.00}$  & 0.53 & (c)   \\
KIC 6225718  & $1.26\pm0.03$                   & $0.08 \pm0.04$                  & 0.60 & (a)  & KIC 8057661     & $9.52\substack{+0.02 \\ -0.03}$ & $1.76\substack{+0.01 \\ -0.01}$  & 0.18 & (c)   \\
KIC 10454113 & $1.27\pm0.02$                   & $0.08 \pm0.03$                  & 0.59 & (a)  & KIC 8255796     & $5.73\substack{+0.10 \\ -0.03}$ & $0.49\substack{+ 0.10 \\ -0.00}$ & 0.02 & (c)   \\
KIC 5184732  & $1.20\pm0.01$                   & $0.06 \pm0.001$                 & 0.26 & (a)  & KIC 8381949     & $6.27\substack{+0.79 \\ -0.00}$ & $1.53\substack{+ 0.07 \\ -0.10}$ & 0.57 & (c)   \\
KIC 12009504 & $1.20\pm0.01$                   & $0.07 \pm0.02$                  & 0.21 & (a)  & KIC 8459899     & $3.36\substack{+0.00 \\ -0.03}$ & $0.84\substack{+ 0.03 \\ -0.00}$ & 0.34 & (c)   \\
KIC 7206837  & $1.44\pm0.04$                   & $0.17 \pm0.03$                  & 0.41 & (a)  & KIC 8714886     & $6.49\substack{+0.00 \\ -0.08}$ & $0.92\substack{+ 0.01 \\ -0.00}$ & 0.28 & (c)   \\
KIC 12258514 & $1.24\pm0.02$                   & $0.09 \pm0.01$                  & 0.07 & (a)  & KIC 8766405     & $3.49\substack{+0.00 \\ -0.23}$ & $0.55\substack{+ 0.03 \\ -0.00}$ & 0.22 & (c)   \\
KIC 7510397  & $1.36\pm0.04$                   & $0.11 \pm0.03$                  & 0.09 & (a)  & KIC 9020774     & $3.41\substack{+0.21 \\ -0.28}$ & $0.70\substack{+ 0.07 \\ -0.17}$ & 0.78 & (c)   \\
KIC 8228742  & $1.33\pm0.05$                   & $0.11 \pm0.01$                  & 0.06 & (a)  & KIC 9715425     & $4.72\substack{+0.45 \\ -0.20}$ & $0.91\substack{+ 0.01 \\ -0.06}$ & 0.48 & (c)   \\
KIC 2710594  & $1.50\substack{+0.13 \\ -0.09}$ & $0.16\substack{+0.01 \\ -0.01}$ & 0.67 & (b)  & KIC 10526294    & $3.64\substack{+0.10 \\ -0.21}$ & $0.41\substack{+0.01 \\ -0.06}$  & 0.23 & (c)   \\
KIC 3448365  & $1.49\substack{+0.01 \\ -0.07}$ & $0.12\substack{+0.01 \\ -0.01}$ & 0.63 & (b)  & KIC 10536147    & $7.50\substack{+0.07 \\ -0.41}$ & $2.17\substack{+0.06 \\ -0.14}$  & 0.98 & (c)   \\
KIC 4846809  & $1.49\substack{+0.05 \\ -0.02}$ & $0.15\substack{+0.01 \\ -0.01}$ & 0.58 & (b)  & KIC 11360704    & $4.47\substack{+1.54 \\ -0.00}$ & $1.20\substack{+0.00 \\ -0.66}$  & 0.47 & (c)   \\
KIC 5114382  & $1.61\substack{+0.02 \\ -0.14}$ & $0.16\substack{+0.01 \\ -0.01}$ & 0.36 & (b)  & KIC 11971405    & $3.67\substack{+0.21 \\ -0.72}$ & $0.62\substack{+0.11 \\ -0.03}$  & 0.46 & (c)   \\
KIC 5522154  & $1.72\substack{+0.28 \\ -0.22}$ & $0.19\substack{+0.03 \\ -0.02}$ & 0.97 & (b)  & KIC 12258330    & $3.53\substack{+0.04 \\ -0.13}$ & $0.69\substack{+0.03 \\ -0.00}$  & 0.68 & (c)   \\
KIC 5708550  & $1.99\substack{+0.01 \\ -0.61}$ & $0.20\substack{+0.01 \\ -0.06}$ & 0.98 & (b)  & KIC 4930889  A  & $4.89\substack{+1.49 \\ -1.09}$ & 0.54                             & 0.07 & (d)   \\
KIC 5788623  & $1.47\substack{+0.01 \\ -0.05}$ & $0.12\substack{+0.01 \\ -0.01}$ & 0.70 & (b)  & KIC 4930889  B  & $3.47\substack{+1.40 \\ -0.57}$ & 0.60                             & 0.68 & (d)   \\
KIC 6468146  & $1.59\substack{+0.01 \\ -0.04}$ & $0.17\substack{+0.01 \\ -0.01}$ & 0.61 & (b)  & KIC 6352430  A  & $3.23\substack{+1.98 \\ -0.56}$ & 0.56                             & 0.59 & (d)   \\
KIC 6468987  & $1.57\substack{+0.13 \\ -0.16}$ & $0.20\substack{+0.02 \\ -0.02}$ & 0.74 & (b)  & KIC 6352430  B  & $1.34\substack{+0.68 \\ -1.14}$ & 0.08                             & 0.77 & (d)   \\
KIC 6678174  & $1.77\substack{+0.23 \\ -0.22}$ & $0.18\substack{+0.02 \\ -0.02}$ & 0.20 & (b)  & KIC 10080943 A  & $1.99\substack{+0.37 \\ -0.49}$ & 0.15                             & 0.08 & (d)   \\
KIC 6935014  & $1.45\substack{+0.03 \\ -0.03}$ & $0.16\substack{+0.01 \\ -0.01}$ & 0.86 & (b)  & KIC 10080943 B  & $1.85\substack{+0.41 \\ -0.45}$ & 0.17                             & 0.30 & (d)   \\
KIC 6953103  & $1.88\substack{+0.12 \\ -0.19}$ & $0.25\substack{+0.02 \\ -0.03}$ & 0.48 & (b)  & V578 Mon A      & $14.55\pm0.09$                  & $5.12\pm0.44$                    & 0.82 & (e)   \\
KIC 7023122  & $1.51\substack{+0.00 \\ -0.04}$ & $0.16\substack{+0.01 \\ -0.01}$ & 0.96 & (b)  & V578 Mon B      & $10.30\pm0.06$                  & $3.21\pm0.37$                    & 0.89 & (e)   \\
KIC 7365537  & $1.70\substack{+0.30 \\ -0.26}$ & $0.18\substack{+0.03 \\ -0.03}$ & 0.97 & (b)  & V453 Cyg A      & $14.95\pm0.35$                  & $4.81\pm0.35$                    & 0.41 & (e)   \\
KIC 7380501  & $1.72\substack{+0.00 \\ -0.35}$ & $0.14\substack{+0.01 \\ -0.03}$ & 0.19 & (b)  & V453 Cyg B      & $11.91\pm0.57$                  & $3.66\pm0.53$                    & 0.66 & (e)   \\
KIC 7434470  & $1.48\substack{+0.01 \\ -0.16}$ & $0.17\substack{+0.01 \\ -0.02}$ & 0.64 & (b)  & V478 Cyg A      & $16.74\pm0.73$                  & $6.04\pm0.85$                    & 0.58 & (e)   \\
KIC 7583663  & $1.47\substack{+0.27 \\ -0.16}$ & $0.16\substack{+0.03 \\ -0.02}$ & 0.64 & (b)  & V478 Cyg B      & $16.39\pm0.71$                  & $5.97\pm0.84$                    & 0.57 & (e)   \\
KIC 7939065  & $1.48\substack{+0.02 \\ -0.06}$ & $0.14\substack{+0.00 \\ -0.01}$ & 0.66 & (b)  & AH Cep A        & $16.26\pm0.18$                  & $6.20\pm0.62$                    & 0.74 & (e)   \\
KIC 8364249  & $1.52\substack{+0.23 \\ -0.05}$ & $0.17\substack{+0.03 \\ -0.01}$ & 0.66 & (b)  & AH Cep B        & $14.86\pm0.70$                  & $5.22\pm0.77$                    & 0.70 & (e)   \\
KIC 8375138  & $1.47\substack{+0.05 \\ -0.07}$ & $0.15\substack{+0.01 \\ -0.01}$ & 0.69 & (b)  & V346 Cen A      & $13.76\pm0.40$                  & $4.14\pm0.35$                    & 0.36 & (e)   \\
KIC 8645874  & $1.52\substack{+0.02 \\ -0.01}$ & $0.14\substack{+0.01 \\ -0.01}$ & 0.88 & (b)  & V346 Cen B      & $8.40\pm0.10$                   & $2.04\pm0.35$                    & 0.73 & (e)   \\
KIC 8836473  & $1.48\substack{+0.11 \\ -0.01}$ & $0.14\substack{+0.01 \\ -0.01}$ & 0.34 & (b)  & V573 Car A      & $16.62\pm0.45$                  & $6.65\pm0.45$                    & 0.87 & (e)   \\
KIC 9480469  & $1.49\substack{+0.05 \\ -0.11}$ & $0.14\substack{+0.01 \\ -0.01}$ & 0.82 & (b)  & V573 Car B      & $12.49\pm0.17$                  & $4.37\pm0.32$                    & 0.95 & (e)   \\
KIC 9595743  & $1.42\substack{+0.03 \\ -0.04}$ & $0.09\substack{+0.01 \\ -0.01}$ & 0.94 & (b)  & V1032 Sco A     & $18.97\pm0.77$                  & $7.44\pm0.89$                    & 0.60 & (e)   \\
KIC 9751996  & $1.74\substack{+0.01 \\ -0.02}$ & $0.17\substack{+0.01 \\ -0.01}$ & 0.29 & (b)  & V1032 Sco B     & $10.47\pm0.24$                  & $3.29\pm0.24$                    & 0.87 & (e)   \\
KIC 10467146 & $1.77\substack{+0.02 \\ -0.27}$ & $0.16\substack{+0.01 \\ -0.02}$ & 0.22 & (b)  & V380 Cyg A      & $15.00\pm0.50$                  & $3.85\pm0.15$                    & 0.00 & (e)   \\
KIC 11080103 & $1.53\substack{+0.02 \\ -0.09}$ & $0.16\substack{+0.01 \\ -0.01}$ & 0.76 & (b)  & V380 Cyg B      & $8.32\pm0.75$                   & $2.26\pm0.59$                    & 0.78 & (e)   \\
KIC 11099031 & $1.57\substack{+0.03 \\ -0.21}$ & $0.17\substack{+0.01 \\ -0.02}$ & 0.41 & (b)  & CW Cep A        & $13.01\pm0.07$                  & $4.38\pm0.62$                    & 0.77 & (e)   \\
KIC 11294808 & $1.45\substack{+0.20 \\ -0.04}$ & $0.13\substack{+0.02 \\ -0.00}$ & 0.27 & (b)  & CW Cep B        & $11.95\pm0.09$                  & $3.86\pm0.48$                    & 0.79 & (e)   \\
KIC 11456474 & $1.49\substack{+0.14 \\ -0.13}$ & $0.13\substack{+0.01 \\ -0.01}$ & 0.25 & (b)  & U Oph A         & $5.10\pm0.05$                   & $1.01\pm0.11$                    & 0.65 & (e)   \\
KIC 11721304 & $1.43\substack{+0.04 \\ -0.03}$ & $0.15\substack{+0.01 \\ -0.01}$ & 0.89 & (b)  & U Oph B         & $4.60\pm0.05$                   & $0.93\pm0.08$                    & 0.74 & (e)   \\
KIC 11754232 & $1.53\substack{+0.02 \\ -0.04}$ & $0.15\substack{+0.01 \\ -0.01}$ & 0.90 & (b)  & V621 Per A      & $11.81\pm0.30$                  & $2.80\pm0.21$                    & 0.84 & (e)   \\
KIC 11907454 & $1.47\substack{+0.01 \\ -0.10}$ & $0.13\substack{+0.01 \\ -0.01}$ & 0.83 & (b)  & HD165246 B      & $3.8\substack{+0.4\\-0.5}$      & $0.42\substack{+0.4\\- 0.2}$     & 1.00 & (f) \\
KIC 11826272 & $1.52\substack{+0.04 \\ -0.07}$ & $0.16\substack{+0.01 \\ -0.01}$ & 0.51 & (b)  & HD165246 A      & $23.7\substack{+1.1\\-1.4}$     & $10.2\substack{+1.3\\- 1.6}$     & 0.76 & (f) \\
KIC 11917550 & $1.51\substack{+0.00 \\ -0.10}$ & $0.16\substack{+0.01 \\ -0.01}$ & 0.67 & (b)  & 2052 Oph        & $8.9\pm0.7$                     & $1.70\pm0.2$                     & 0.41 & (g)   \\
KIC 11920505 & $1.45\substack{+0.03 \\ -0.01}$ & $0.11\substack{+0.01 \\ -0.01}$ & 0.88 & (b)  & HD~43317        & $5.8\substack{+0.1\\-0.2}$      & $1.33\pm0.05$                    & 0.76 & (h)   \\
KIC 12066947 & $1.49\substack{+0.02 \\ -0.01}$ & $0.14\substack{+0.01 \\ -0.01}$ & 0.65 & (b)  & HIC 12258514    & $1.20\pm0.05$                   & $0.05\pm0.01$                    &      & (i)   \\
KIC 3240411  & $5.34\substack{+0.31 \\ -0.07}$ & $1.56\substack{+0.00 \\ -0.02}$ & 0.68 & (c)  & KIC 8228742     & $1.25\pm0.08$                   & $0.06\pm0.02$                    &      & (i)   \\
KIC 3459297  & $3.70\substack{+0.00 \\ -0.18}$ & $0.24\substack{+0.01 \\ -0.01}$ & 0.05 & (c)  & KIC 7510397     & $1.22\pm0.06$                   & $0.07\pm0.03$                    &      & (i)   \\
KIC 3865742  & $5.66\substack{+0.00 \\ -0.08}$ & $1.05\substack{+0.03 \\ -0.00}$ & 0.22 & (c)  & KIC 10162436    & $1.30\pm0.06$                   & $0.09\pm0.02$                    &      & (i)   \\
KIC 4936089  & $3.66\substack{+0.01 \\ -0.00}$ & $0.43\substack{+0.00 \\ -0.00}$ & 0.35 & (c)  & KIC 5773345     & $1.49\pm0.09$                   & $0.15\pm0.01$                    &      & (i)   \\
KIC 4939281  & $5.38\substack{+1.18 \\ -0.34}$ & $1.31\substack{+0.01 \\ -0.27}$ & 0.48 & (c)  & KIC 3456181     & $1.39\pm0.07$                   & $0.11\pm0.03$                    &      & (i)   \\
KIC 5309849  & $5.67\substack{+0.01 \\ -0.00}$ & $0.75\substack{+0.00 \\ -0.00}$ & 0.15 & (c)  & KIC 8179536     & $1.22\pm0.09$                   & $0.03\pm0.03$                    &      & (i)   \\
KIC 5941844  & $3.75\substack{+0.03 \\ -0.02}$ & $0.90\substack{+0.04 \\ -0.27}$ & 0.94 & (c)  & $\beta$ Cma     & $13.5\pm0.5$                    & $3.5\pm0.3$                      & 0.18 & (j)   \\
KIC 6462033  & $7.45\substack{+0.13 \\ -0.00}$ & $1.08\substack{+0.05 \\ -0.00}$ & 0.17 & (c)  & $\beta$ Cru     & $14.5\pm0.5$                    & $4.13\pm0.5$                     & 0.27 & (k)   \\
KIC 6780397  & $6.29\substack{+0.01 \\ -0.00}$ & $0.55\substack{+0.00 \\ -0.00}$ & 0.05 & (c)  & KIC~9850387     & $1.66\pm0.01$                   & $0.16\pm0.01$                    & 0.24 & (l)  \\
KIC 7630417  & $6.92\substack{+0.01 \\ -0.00}$ & $0.91\substack{+0.00 \\ -0.00}$ & 0.15 & (c)  &                 &                                 &                                  &      &      \\

\hline
\end{tabular}
\tablebib{ 
(a)~\citealt{Angelou2020} ; (b)~\citealt{Mombarg2021} ; (c)~\citealt{Pedersen2021} ;
(d)~\citealt{Johnston2019a}; (e)~\citealt{Tkachenko2020}; (f)~\citealt{Johnston2021};
(g)~\citealt{Briquet2011}; (h)~\citealt{Buysschaert2018b}; (i)~\citealt{Viani2020};
(j)~\citealt{Mazumdar2006}; (k)~Cotton et al. {\it in review}; (l)~\citealt{Sekaran2021}
}

\end{table*}

\end{appendix}

\end{document}